\documentclass[conference]{IEEEtran}
\IEEEoverridecommandlockouts
\usepackage{cite}
\usepackage{amsmath,amssymb,amsfonts}
\usepackage{url}
\usepackage{graphicx}
\usepackage{textcomp}
\usepackage{xcolor}
\def\BibTeX{{\rm B\kern-.05em{\sc i\kern-.025em b}\kern-.08em
    T\kern-.1667em\lower.7ex\hbox{E}\kern-.125emX}}
    
\usepackage[printonlyused]{acronym}

\usepackage{comment}

\usepackage{subcaption}

\ifCLASSINFOpdf
\graphicspath{{pdf/}{jpeg/}{eps/}}
\DeclareGraphicsExtensions{.pdf,.jpeg,.eps}
\else

\DeclareGraphicsExtensions{.pdf,.jpeg,.png,.eps}
\else
\usepackage[dvips]{graphicx}
\fi

\usepackage{float}
\usepackage{caption}

\begin{document}

\title{5G Simulation-Based Experimentation Framework for Vertical Performance Assessment\\
\thanks{\textcopyright 2020 IEEE. Personal use of this material is permitted. Permission
	from IEEE must be obtained for all other uses, in any current or future
	media, including reprinting/republishing this material for advertising or
	promotional purposes, creating new collective works, for resale or
	redistribution to servers or lists, or reuse of any copyrighted
	component of this work in other works.}
}

\author{\IEEEauthorblockN{Orestes Manzanilla-Salazar}
\IEEEauthorblockA{\textit{dept. of Electrical Engineering} \\
\textit{Ecole Polytechnique}\\
Montreal, Canada \\
orestes.anzanilla@polymtl.ca}
\and
\IEEEauthorblockN{Victor Boutin}
\IEEEauthorblockA{\textit{dept. of Electrical Engineering} \\
\textit{Ecole Polytechnique}\\
Montreal, Canada \\
victor.boutin@polymtl.ca}
\and
\IEEEauthorblockN{Hakim Mellah}
\IEEEauthorblockA{\textit{dept. of Electrical Engineering} \\
\textit{Ecole Polytechnique}\\
Montreal, Canada \\
hakim.mellah@polymtl.ca}
\and
\IEEEauthorblockN{Constant Wett\'e}
\IEEEauthorblockA{\textit{Ericsson} \\
Montreal, Canada \\
constant.wette.tchouati@ericsson.com}
\and
\IEEEauthorblockN{Brunilde Sans\`o}
\IEEEauthorblockA{\textit{dept. of Electrical Engineering} \\
\textit{Ecole Polytechnique}\\
Montreal, Canada \\
brunilde.sanso@polymtl.ca}

}

\maketitle

\begin{abstract}
5G is being designed as a common platform where multiple vertical applications will be able to co-exist and grow in a seamless manner. 
The diversity of the vertical requirements as well as the particular features of the 5G network itself, make it a real challenge to be able to assess or predict applications performance for those verticals.
In this paper we motivate the fact that because of the very nature of verticals and 5G infrastructure, measurements alone will not be sufficient for application performance monitoring and prediction. 
We then propose a comprehensive framework to integrate those measurements in a simulation setting, presenting the key features and current roadblocks for an end-to-end wide-spread implementation.

\end{abstract}


\section{Introduction}

Smart-cities deployment  enable emerging applications and services in many verticals such as intelligent transportation systems (ITS) and utility systems \cite{yadav18,khurpade18,rusti18}.
Current 4G mobile technologies such as Long-Term Evolution (LTE) have been used as the telecommunication infrastructure for IoT applications. However, 4G technologies are not suitable for massive IoT applications \cite{akpakwu18}.
5G is therefore the technology of choice for next generation enhanced Mobile Broadband (eMBB), ultra-reliable low latency communications (URLLC) and massive machine-type communications (mMTC), commonly refered to as \textit{5G triangle} \cite{aldulaimi18,vu17,she19,ji18,ashraf17}.  

Many smart-city vertical applications are geographically distributed in nature. They depend not only on the immediate access features, but also on the behaviour of other, often remote,  applications that share the same 5G infrastructure. Another important role for a successful application deployment is played by the flexibility features of the 5G, such as the mapping of slices, their physical routing and the access adaptive functions (cell zooming, antenna tilting, on-off operation, etc.)
Alongside the diversity of the applications, their geographically distributed nature and the adaptability of the infrastructure, comes the uncertain 5G propagation behavior in the mmWave spectrum \cite{khalily18,alsamman16}. 

As a result, 5G operators will be facing the problem of having to deliver highly reliable applications, often mission critical in nature, in a largely distributed system with high variability both in terms of traffic and signal reception. 
It is clear then that network and application performance assessment can no longer be confined to local measurements.
Operators will need comprehensive tools to be able to detect, infer or predict how applications are doing.
Moreover, they will also need  to be able to determine the impact that adaptive features changes bring to network and application performance. 

In our view,  large-scale measurement campaigns will not be enough to capture network adaptability in such a complex technological setting. 
We propose, instead  a framework in which 
a geographically distributed network simulator alongside measurement campaigns enable machine learning algorithms to assess application behaviour, evaluate the impact of operational changes, detect network or applications anomalies and predict performance enhancement or deterioration. 

Network simulators are generally small scale and can rarely capture the end-to-end performance of a city network. 
A notable exception is the city-wide 4G simulator developed by our group  \cite{malandra17,malandra17a,malandra18} where end-to-end smart-city applications performance can be evaluated and failures can be detected \cite{manzanilla19}. 
However, extending the same ideas to the 5G context presents some interesting challenges that deal with the very nature of the new applications and infrastructure, in particular scalability and propagation modelling.

In the remainder of the paper, the following material is covered. In Section \ref{state-of-the-art} , we review the State of the art on 5G physical layer challenges, propagation models, current 5G simulators and particular 5G and vertical features. 
The proposed framework is introduced in Section \ref{framework}.
Concluding remarks are presented in Section \ref{conclusions}

\section{State of the Art} \label{state-of-the-art}

\subsection{5G physical layer challenges and propagation models}

To increase wireless communications capacity, 5G relies on 
previously unused frequencies in cellular networks. These 
frequencies, are the upper centimetre-wave band (3-30 GHz) and 
the mmWave band (30-300 GHz) \cite{sulyman16}. 
Higher frequencies change the 
way propagation channels must be modelled, in part due to 
 higher path loss, but also because they enable the use of 
new technologies. 

By the Friss law, we know that the path loss in free space is 
proportional to the square of the frequency, which makes the 
mmWaves isotropic range lower than the frequencies used in 4G. The trends of 
4G Networks is densification due to 
lack of capacity. So even if the 5G requires a higher density, 
the magnitude of cell radius needed is familiar to 4G \cite{rangan14}.

With antennas range decreasing, a need to agglomerate emerges. To achieve this goal, 
multiple transmission and reception points (TRP) can be grouped to form a single cell 
\cite{liu18}. This single cell is an abstraction for the network layer (L3) making 
handover between TRP of a same cell handled at lower layer (L1 and L2), which would 
be otherwise handled up to the network layer.

The main issue for channel modelling is the high penetration loss.
Attenuation for common building surfaces such as tinted glass 
and brick is 40.1 dB for the first and 28.3 dB for the second
 \cite{rappaport13}. 
With this much path loss, the chance of outage is significantly higher.
Obstacles such as building, vegetation and people have 
important impact on the received power. 
Outages are a main concern for designing cellular networks \cite{sun18, akdeniz14}.
To 
mitigate this problem, dual-connectivity with legacy 4G could be used to stay connected
\cite{mezzavilla18}, with reduced performance. The fast channel changes should also be 
considered in the scheduling, as mentioned by \cite{ford17}. They not only affect the MAC
layer (L2), but the TCP protocol, which cannot manage fast variation of throughput
\cite{mezzavilla18}.
Also, outdoor to indoor communication will be quite limited. 
On the flip side, the high penetration loss makes it possible 
to isolate outdoor and indoor networks \cite{rappaport13}.

As \cite{maccartney13} mentioned, the channel is highly dependant 
on the environment, even in similar contexts (e.g: urban micro cell).
This makes statistical/empirical models hard to transpose without 
actually taking field data of the region to be modelled.
In the case of deterministic models, the higher the frequency, the more 
detailed must be the environment \cite{shafi18}. Furthermore, as the environment model 
resolution increases, the computation time also increases making them unfit
for large scale modelling.

Because the minimal size of an antenna is proportional to the wavelength,
massive multiple input, multiple output (MIMO) antenna technology can be used to its full potential. This 
technology is crucial for 5G. In conjunction with beamforming, 
path loss gain can be mitigated by antennas gain \cite{akdeniz14}, without 
increasing the actual antenna size. On the 
other hand, it adds a degree of complexity to properly align beams, in particular 
during connections. To establish connections, a process of beam-sweeping is used to 
synchronize the transmitter and receiver \cite{liu18}.

With higher frequencies comes higher path loss in rain and in the atmosphere 
in general, but these attenuations can be considered negligible compared 
to other path loss in cellular network. Indeed, considering a rainfall of 
1 inch/hour, for an 
RX at 200 m (edge of the cell), the attenuation is 1.4 dB at 28 GHz. As for
the atmospheric attenuation, most frequencies between 10 and 100 GHz have an
attenuation of less than 1 dB/km \cite{rappaport13}, the only exception 
begin the frequency 
band around 60 GHz, which is researched for WLAN \cite{akdeniz14}.

The frequencies used for 4G and before are now well studied 
and tested. Multiple models have been used to estimate the path loss like 
the Hata model and its extension COST Hata. The Hata model, which was developed 
in 1980, proposes a path loss equation for a frequency range of 100-1500 MHz, a 
distance of 1-20km and antenna height of 30-200m \cite{hata80}. It was later 
extended (COST-Hata) to cover frequencies up 2 GHz. These models 
are clearly not fit for 5G modelling, considering the high frequency, low distance 
and low height of the antennas.

Path loss models only consider the losses due to distance and shadowing. To consider 
the effect of multipath (small scale fading), statistical 
distributions such as Rayleigh, for non-line-of-sight (NLOS), 
or Rice, for line-of-sight (LOS), have been used to
model the received gain\cite{sun18}.
By combining the path loss models with the multipath statistical distribution 
gain/loss, a satisfying estimation of the propagation can be established.

As for the 5G, mmWave are already used in some applications such as cellular 
backhaul and satellite communications \cite{rangan14}, but not in a mobile 
NLOS context. Also, research has been done 
on 60GHz in indoor environment for WLAN purpose. For outdoor urban uses, many data 
collection campaigns published results \cite{maccartney13, nguyen16}, such as 
the New York city data collection by the NYU Polytechnic where a single unknown 
variable model is developed \cite{rappaport15}:

\begin{equation}
PL(d)[dB] = PL(d_0) + 10nlog_{10}\left(\frac{d}{d_0} \right) + X_\sigma \label{eq:CI_model}
\end{equation}

where:
\begin{itemize}
	\item $d_0$ is an arbitrary distance
	\item $PL(d_0)$ is the free space path loss (FSPL) at $d_0$ and depends on the frequency
	\item $n$ is the path loss exponent (PLE)
	\item $d$ is the evaluated distance
	\item $X_\sigma$ is random shadowing variable following a Gaussian distribution
\end{itemize}

The model is based on urban microcells experiments based in
New York city and Austin.

Another, approach consists 
of using two variables, $\alpha $ and $\beta$ (AB model), to build the 
model instead of a single variable ($n$) like the previous model. The 
$\alpha$ parameter is the equivalent of the FSPL, but it is not based on physical 
characteristics of the channel, and $\beta$ is the equivalent of $n$ 
\cite{rappaport15, 3GPP20a} which represent the path loss exponent. 
In essence, the model is a linear regression 
fit with added noise. Because the former takes into account the
frequency, it is more flexible an can be used on a
wider range of frequencies, compared to the AB model, which is highly dependant 
of measurement settings, but gives a more exact fitting.

Other studies such as \cite{nguyen16} concentrate on different environments, 
in this specific case, urban macrocells. The AB model was used and the 
results were that, for LOS, the 3D distance between the TX and RX was the 
main source of path loss. On the other hand, for NLOS, the antenna height 
had a major effect on the result. By placing the antenna below the height 
of buildings (15m), the path loss was 23.4 dB higher than a 54m TX.

Two main models have been generalized based on the previous models: NYUSIM and 
3GPP. Even though both
use a similar approach, there are many variation between the models. In 
\cite{sun18}, the two models are compared with each other.

\subsection{5G Simulators and measures}

To deploy new technologies, simulators are crucial. Because of the 
differences between 4G and 5G networks and in particular the channel model,
as mention in the earlier sections,
4G simulators can't be directly used to model 5G networks.

Some low level simulators where made from the ground up, such as the NYUSIM.
The NYUSIM is an open-source 5G channel simulator written in MATLAB based on 
the close-in free space reference distance (CI) propagation model \cite{sun17}.

Others based their simulator on ns-3, such as \cite{mezzavilla18} in which an 
extension for ns is described. The extension implements a custom PHY and MAC layer
as well as a custom channel model. It is used in conjunction with the ns-3 LTE 
simulator (LENA) to simulate end-to-end traffic. In \cite{5gwise19} a multi-part 
simulator is presented with each part having a different level of abstraction. 
The higher level part, K-SimSys, used for end-to-end simulation, is also based 
on ns-3.

All the previous mentioned simulators present an in depth simulation of the channel 
and/or a detailed MAC layer. This level of details makes the simulation 
precise, but at the same time, computer intensive. These simulation approaches
are unsuitable for large scale simulation and traffic analysis with mobility.

Most of these simulatiors are based on purely theoretical models or LTE models. 
An exception is the NYUSIM, which uses the CI model, based on \eqref{eq:CI_model}
 which was developed with data obtained during data collection campaigns in 
 Manhattan and Austin where steerable 
horn antennas with 24.5 dBi were used both as transmitters and receivers
\cite{maccartney13}. Even in this case, the experience made abstraction of many
5G aspects to concentrate only on mmWaves. As mentionned in \cite{Foegelle19}, 
 testings in 5G environment bring complications, in particular concerning beam 
 forming and steering.

\section{Simulation-based vertical assessment} \label{framework}

\subsection{Smart City Verticals}

A \emph{vertical} in a smart city is a set of technologies whose goal is to satisfy a type of human needs. Examples of verticals are the \acp{ITS}, smart waste management systems and the smart grids. We define three layers in a vertical:

\begin{enumerate}
	\item \emph{Human interaction layer:} this is a layer where humans (users, owners, administrators, clients or operators) interact with a system of the vertical.
	\item \emph{Applications layer:} a vertical may have one or more applications or functionalities. In smart-grid applications for example, some applications are permanently detecting the topology of the power network, while other applications might be in charge of protection.
	\item \emph{Machines layer:} each application's task may depend on one or more types of devices. These may be geographically distributed along the city, have different traffic profiles and use different telecommunication technologies.
\end{enumerate}

The 5G infrastructure might be used in any of the vertical layers, from human interaction to the communications between devices.

To assess the performance of a vertical in a smart city, we define three types of \acp{KPI}:

\begin{enumerate}
	\item \emph{User KPI}: this kind of indicator is designed to assess the experience of a user. For example, in the case of an \ac{ITS} application, error on the estimated arrival time of a bus might be considered a User KPI. These indicators may involve time scales in the order of several seconds or minutes.
	\item \emph{Network KPI:} these are the performance measurements available to the network infrastructure operator (delay, jitter, packet loss, throughput, etc).  The relevant time scales for these indicators are typically in the order of milliseconds.
	\item \emph{Vertical KPI:} at a system-level, the relevant indicators will be related to the extent to which the system is being able to achieve its goal. In the example of \ac{ITS} applications, the percentage of users of the bus application that have a ``bad experience'' in a specific day is relevant. The time scale may be in the order of days.
\end{enumerate}

The network \acp{KPI} may be used by an operator to infer problems in user \acp{KPI}, in cases where specific performance thresholds have been identified. Vertical \acp{KPI}, however, might be difficult to infer to the operator. User and vertical \acp{KPI} might have to be inferred by using simulation and \ac{ML} techniques, or considering user surveys. In many applications, the network requirements and \acp{KPI} are extremely specialized. As an example, synchrophasor devices in a smart grid, have requirements that depend on their reporting rate.

\subsection{Multi-vertical large-scale simulation approach}

The strategy we propose to deal with the immense computational burden that would be involved in implementing a city-wide 5G simulator is to avoid simulating in detail (physical layer, propagation and protocols) at the city scale. This relevant specially in presence of verticals with massive \ac{IoT} applications. The approach is to create a simulator based on cell-level simulators using ``realistic'' random generation of \acp{KPI} for each packet.

The biggest issue with simulating without considering physical and protocols details at millisecond level for each packet, is potential lack of realism. In order to obtain ``realistic'' \acp{KPI} the statistical distributions used to generate the data should be similar to the variability patterns these \acp{KPI} really have, given the \emph{conditions} in the 5G cell, such as number of devices and traffic patterns.

In order to estimate the \acp{PDF} of the \acp{KPI} of the packets generated in each cell, our simulator uses several components which will now be described.

\subsubsection{Detailed cell-level 5G simulator}
\begin{figure}[h]
	\begin{subfigure}[t]{.2\textwidth}
		\centering
		\includegraphics[width=0.8in]{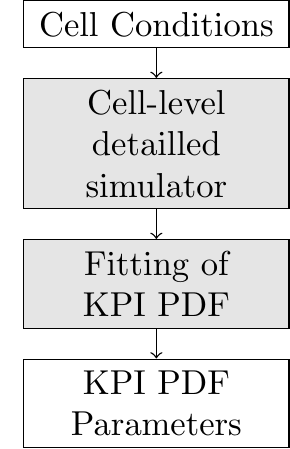}
		\centering	
		\caption{Detailed simulator.}
		\label{figure:abst1}
	\end{subfigure}
	\begin{subfigure}[t]{.2\textwidth}
		\centering
		\includegraphics[width=0.8in]{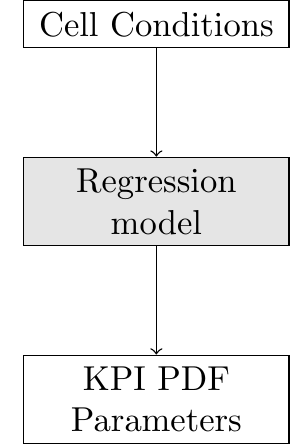}
		\centering	
		\caption{Regression model.}
		\label{figure:abst2}
\end{subfigure}
	\caption{Detailled simulator abstraction (data in white, process in grey).}
\end{figure}

A detailed simulator, considering the modelling of the physical layer, and 5G protocols, needs to be implemented, considering only one cell. This simulator should be run in a wide multidimensional parameter-sweep, gathering packet-level \acp{KPI} in each experimental condition. In each one, a \ac{PDF} should be fitted, to be later used to describe the variability of the \acp{KPI} under these cell conditions (see Fig. \ref{figure:abst1}).

\subsubsection{Random KPI generator}

Having pairs of sets of cell conditions and the parameters of the \acp{PDF} of the resulting \acp{KPI} allows the use of regression techniques to estimate the \acp{PDF} parameters (see Fig. \ref{figure:abst2}). This is an essential step to be able to have a realistic random generation of the \acp{KPI}. The random generator will use these parameters to evaluate for each packet in the simulation the inverse of the cumulative probability density function (CPDF), in a rather standard way: a random value distributed uniformly in the $[0,1]$ range has to be generated and evaluated in the CPDF to obtain values whose distribution matches the \ac{PDF} with the same parameters. The similitude between the generated data variability and that of those generated by the detailed simulator will depend on the quality of the regression model.

\subsubsection{Urban activity simulator}

In a realistic city model, conditions in each cell are variable (number of sources of traffic and their traffic profiles), specially if we consider devices with mobility, which is essential when modelling verticals like the \ac{ITS}. The question arises: \emph{how to generate a dynamic system, when the data generation assumes constant conditions?}. Our proposed answer is the use of a \emph{one-way} urban activity simulator.

By urban activity, we refer to the activities undertaken in the city, which may affect the generation of packets, or the location of the sources of traffic. If we consider cars connected to the 5G network, the urban activity simulator will model each car in the city in their trajectory. The design of the simulation models needs to be focused on the behaviors that can be associated to changes in cell conditions, for example, the events of handover. We refer to this simulator as a one-way process, as it is assumed to be unaffected by the telecom system. There is no feedback between the urban activities and the telecom system events.

The output of the urban simulator should consist on a list of time intervals of constant conditions for each cell in the city. Each interval should be accompanied with the cell conditions present along the time interval. These, in general, may be different across cells, as during one interval one cell may have a constant condition, while another one experiences several events in which its conditions change.

The urban simulator is important as its output allows the integration of the cell-level random-generation of the \acp{KPI} into a city-level simulation, considering in a coherent way the condition changes of each cell. If a car's connection experiences a handover from one cell to another, this involves an increase in the number of traffic sources in the destination cell, and a decrease in its origin cell. Both cells change in the same event, in different ways.

Parameters based on real information and statistics related to the urban activity should be used to run the urban simulator. As an example, real bus schedules and trajectories can be used to make a simulation of an application related to the public transportation system.

During the implementation of the urban activity simulator, some slack can be used to decide when the condition changes in a cell are big enough to be accounted for. This is a decision that should be addressed during the development of the simulator, depending on the goals of the researcher.

Each interval for each cell, defines the domain in which the \ac{PDF} parameters evaluated for the cell conditions can be used in the random generation of the packets produced during said interval. The number of packets to be generated can be determined using the urban simulation modeling. Let us have 20 cars in a cell during an interval of 5 seconds. Let us have a packet generation rate of one packet per second in each car. Under this situation we would have a total of 100 packets being generated by the \ac{ITS} application. For each one, the \acp{KPI} are generated using the random generator.

\begin{figure*}[t] 
	\centering
	\includegraphics[width=5.0in]{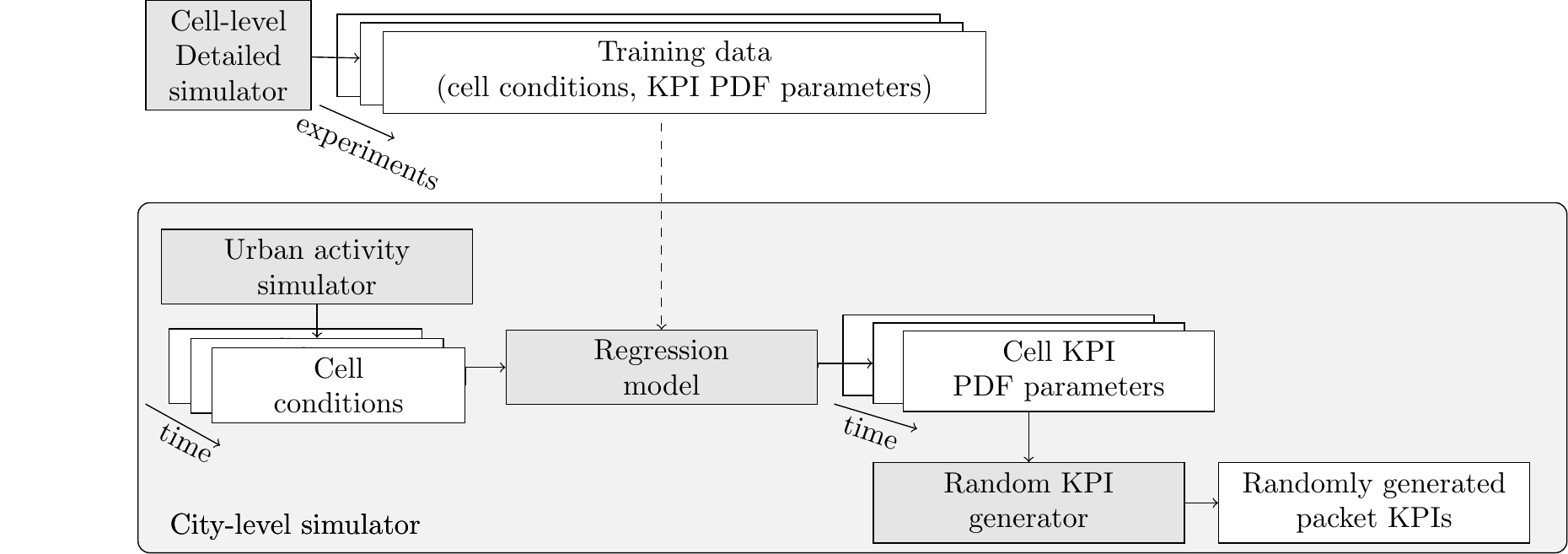}
	\centering	
	\caption{Interplay of city-wide simulator elements.}
	\label{figure:layers}
\end{figure*}

The time in which packets are generated can also be modeled, by assuming a traffic regime (a Poisson process, periodic synchronous or asynchronous generation, etc.). This way we can generate the time-stamps for each packet in the system, if it is an important aspect to take into account in the research.

The city-wide simulation process is depicted in Fig. \ref{figure:layers}. The simulation starts with the urban system, which generates the intervals of constant conditions for each cell. The regression model is used in each interval to obtain the \ac{PDF} parameters required to run the random generator under the conditions of each interval and cell. The generator is run once per each packet generated to assign them their \acp{KPI}. 

\subsubsection{Model validation}

For a simulator to be useful, it is necessary to validate its results. In this framework validation is needed in two aspects: (a) the simulation of urban activity, and (b) of \ac{KPI} generation. The validation of the model of urban activity should be carried out with focus on the way the simulated activity changes cell conditions. The \ac{KPI} generation should be validated by performing measurements in a real 5G base station, and running the detailed simulator under the conditions of the real cell to evaluate how similar they are. How similar these should be, is a modeling decision to be taken by the designer of the specific simulator.

For urban activity, real urban activity should be monitored in an area covered by real 5G cells, and the changes in conditions should resemble those showin in the simulator under similar urban activity simulation parameters. In the case of handovers, for example, the size of the intervals of constant number of devices should be similar to those observed in the real area served by 5G cells.
In any case, meassurements and experimentation with a real 5G infrastructure is an essential need.

\subsection{Simulation-based what-if analysis}

Once such a large-scale city-wide simulator is built, it can be used to pose experiments that may help in the decision making processes involving both from the infrastructure operator point of view, and the ``owner'' of an \ac{IoT} enabled smart-city application. Following a ``what if'' strategy, the analyst would consider the parameters of the experimental scenario, and introduce it into the 5G large-scale simulator, obtaining \acp{KPI} which will need to be analyzed. Depending on the goals of the analyst, adjustments to the infrastructure, the traffic profiles, or the urban activity patterns can be introduced to evaluate the system in a new simulation.

\subsection{Use cases in decision making}

The simulation experimental framework can be used in tasks relevant for operators and ``owners'' of verticals, in the decision making processes involved in the design, adjustment and optimization of their systems, as well the evaluation of hypothetical scenarios to design proper contingencies that may allow their systems to maximize their continuity. We present here some examples of potential uses of the framework:

\begin{enumerate}
	\item \emph{Wireless infrastructure design:} operators may be able to modify the power, orientation, frequency and location of the antennas of their \acp{BS} as well as add or remove \acp{BS}, and study the coverage and quality of the services offered. 
	\item \emph{Contingencies for catastrophic events:} Via simulation, random or deterministic outages or failures assumed as consequences of a large-scale environmental catastrophe in the city can help predict the severity of the event effects, as well as the impact of possible compensation adjustments.
	\item \emph{Flood cyber-attacks:} It is possible to evaluate vulnerability analysis as well as fast-response detection models in face of flooding attacks. These kinds of attacks can be simulated by adding new traffic sources with traffic profiles that designed to stress the system enough to generate degradation on the \acp{KPI} of the \ac{IoT}-enabled application targeted by the attack.
	\item \emph{Failure detection:} By simulating different kinds and profiles of failures, applications \acp{KPI} sensitivity and robustness can be studied. This can be also be helpful in designing compensation strategies.
	\item \emph{Application design:} the ``owners'' of \ac{IoT} applications can evaluate, given a real (or hypothetical) wireless infrastructure, the ways in which network performance may affect the ability of the applications to achieve their goals. The analyst may test different communications strategies and traffic profiles until the application's communications have been adapted to coexist with the assumed congestion levels and infrastructure.
\end{enumerate}

\section{Conclusion} \label{conclusions}
In this paper, we propose a comprehensive and scalable framework blending measurements and simulations for vertical applications performance assessment in a 5G-based smart-city context. This framework would assist the infrastructure stakeholders (owners, operators and users) in monitoring and predicting the performances of the widespread applications using the infrastructure while relieving them from conducting expensive large-scale measurements. Furthermore, the proposed framework would also support the stakeholders in decision making for designing, adjusting and optimizing their systems and applications.

\acrodef{ARMA}{Auto-regressive moving average}
\acrodef{ARIMA}{Auto-regressive integrated moving average}
\acrodef{DMC}{Data Management Center}
\acrodef{EIRP}{Equivalent Isotropic Radiated Power}
\acrodef{HTC}{Human Type Communication}
\acrodef{KPI}{Key Performance Indicator}
\acrodef{IoT}{Internet of Things}
\acrodef{M2M}{Machine-to-Machine}
\acrodef{MTC}{Machine Type Communication}
\acrodef{BS}{Base Station}
\acrodef{CQI}{Channel Quality Indicator}
\acrodef{SON}{Self-Organizing Network}
\acrodef{UE}{User Equipment} 
\acrodef{OSS}{Operation Support System}
\acrodef{RACH}{Random-Access Channel}
\acrodef{RAO}{Random-Access Opportunity}
\acrodef{RB}{Resource Block}
\acrodef{RSRP}{Reference Signals Received Power}
\acrodef{RSRQ}{Reference Signals Received Quality}
\acrodef{SOM}{Self Organizing Map}
\acrodef{SMOTE}{Synthetic Minority Over-Sampling Technique}
\acrodef{SVM}{Support Vector Machine}
\acrodef{QoS}{Quality of Service}
\acrodef{QoE}{Quality of Experience}
\acrodef{microPMU}{Micro-Phasor Measuring Unit}
\acrodef{L-SVM}{Linear Support Vector Machine}
\acrodef{BDT}{Bagged Decision Trees}
\acrodef{PMU}{Phasor Measurement Unit}
\acrodef{IoT}{Internet of Things}
\acrodef{HTC}{Human-Type Communications}
\acrodef{AUC}{Area Under the Curve}
\acrodef{ROC}{Receiver Operating Characteristic}
\acrodef{BS}{Base Station}
\acrodef{FPR}{False Positive Rate}
\acrodef{RBF}{Radial Basis Function}
\acrodef{RSS}{Received Signal Strength}
\acrodef{LTE}{Long-Term Evolution}
\acrodef{3GPP}{$3^{rd}$ Generation Partnership Project}
\acrodef{4G}{$4^{th}$ Generation of broadband cellular network technology}
\acrodef{5G}{$5^{th}$ Generation}
\acrodef{ML}{Machine Learning}
\acrodef{EPC}{Evolved Packet Core}
\acrodef{ITS}{Intelligent Transportation System}
\acrodef{PDF}{Probability Density Function}

\bibliographystyle{IEEEtran}
\bibliography{5g_verticals_framework}{}

\end{document}